# Interfacial properties of black phosphorus/transition metal carbide van der Waals heterostructures

Hao Yuan, Zhenyu Li[*]

Hefei National Laboratory for Physical Sciences at the Microscale, University of Science and Technology of China, Hefei, Anhui 230026, P. R. China

[*]Corresponding author. zyli@ustc.edu.cn

Owing to its outstanding electronic properties, black phosphorus (BP) is considered as a promising material for next-generation optoelectronic devices. In this work, devices based on BP/MXene ($Zr_{n+1}C_nT_2$, T = O, F, OH, n = 1, 2) van der Waals (vdW) heterostructures are designed via first-principles calculations. $Zr_{n+1}C_nT_2$ compositions with appropriate work functions lead to the formation of Ohmic contact with BP in the vertical direction. Low Schottky barriers are found along the lateral direction in BP/$Zr_2CF_2$, BP/$Zr_2CO_2H_2$, BP/$Zr_3C_2F_2$, and BP/$Zr_3C_2O_2H_2$ bilayers, and BP/$Zr_3C_2O_2$ even exhibits Ohmic contact behavior. BP/$Zr_2CO_2$ is a semiconducting heterostructure with type-II band alignment, which facilitates the separation of electron-hole pairs. The band structure of BP/$Zr_2CO_2$ can be effectively tuned via a perpendicular electric field, and BP is predicted to undergo a transition from donor to acceptor at a 0.4 V/Å electric field. The versatile electronic properties of the BP/MXene heterostructures examined in this work highlight their promising potential for applications in electronics.

# 1 Introduction

Two-dimensional (2D) materials are expected to play an important role in electronics, due to their intriguing electronic and optical properties. Monolayer and few-layer black phosphorus (BP) with direct band gap and high carrier mobility are among the most promising channel materials for future nanodevices [1–3]. BP-based field-effect transistors (FETs) with a mobility up to 1000 cm$^2$ V$^{-1}$ s$^{-1}$ and a current on/off ratio in the range of $10^4$–$10^5$ have been successfully fabricated. In these BP FETs, metals such as Au and Ti are directly deposited on the BP surface as electrodes [4]. The strong interaction between BP and the metal electrode can suppress the intrinsic properties of BP and limit the device performance. At the same time, a high contact resistance at the metal-BP interface, dominated by the Schottky barrier, will also limit the carrier injection efficiency [5–7].

A reduction in the Schottky barrier is difficult to achieve when a common metal is used as the electrode, due to the strong Fermi level pinning caused by interfacial states. In this case, a possible way to improve the FET performance involves inserting a thin buffer layer between metal and semiconductor [8–11]. Alternatively, the Fermi level pinning can be suppressed by replacing the common metal used as electrode with a 2D metal. For example, it has been demonstrated that the Schottky barrier becomes tunable or even vanishes when 2D metal electrodes based on 2D metal/NbS$_2$ van der Waals (vdW) heterostructures are used [12].

Previous studies have identified a large number of 2D metallic materials, with graphene as a prominent example [13]. A new family of 2D metals consisting of transition metal carbides/nitrides (MXenes) has recently been discovered. MXenes are fabricated by selective etching of A elements (where A is a group 13 or 14 element such as Al and Si) from MAX phases with general formula $M_{n+1}AX_n$ (n = 1, 2, 3), where M represents an early transition metal and X = C or N. More than twenty MXene materials have been successfully synthesized to date [14–17], and the MXene family is still growing. Due to their active metal surface, MXenes are typically terminated by functional groups such as O, F, or OH [14]. With the exception of several oxygen-terminated MXenes (Ti$_2$CO$_2$, Zr$_2$CO$_2$, and Hf$_2$CO$_2$) and Sc$_2$CT$_2$ (T = O, F, or OH) [18–21], most MXenes are metallic and exhibit excellent electronic conductivity [14]. Therefore, they can be used as electrode materials for future nanoelectronic devices. A low hole Schottky barrier without pinning effects was recently reported for $WSe_2$ FETs based on $Ti_2C(OH)_xF_y$ electrodes [22]. Theoretical studies also predicted that Schottky barrier-free contacts can be realized in 2D semiconductor/MXene heterostructures [23].

Besides their electrode applications, MXenes can also be assembled with other 2D materials to fabricate heterostructures. Various vdW heterostructures based on 2D materials have recently been designed. Owing to the weak interaction between the 2D materials in such structures, their intrinsic properties can be largely preserved. The formation of vdW heterostructures is also an effective method to protect the 2D materials from ambient oxidation. For example, BN/BP/BN sandwich heterostructures have been used to preserve the quality and high carrier mobility of BP [24]. In addition, semiconducting heterostructures with type-II band alignment can enhance the device performance. For instance, the $BP/MoS_2$ vdW *p-n* heterostructure is considered to be a very promising candidate for nano- and optoelectronic applications [25].

In this work, we systematically investigate BP/MXene vdW heterostructures using first-principles calculations. An Ohmic contact along both the vertical and lateral directions is predicted in the case of $BP/Zr_3C_2O_2$, whereas an Ohmic contact in the vertical direction and a low Schottky barrier in the lateral direction are found in $BP/Zr_2CF_2$, $BP/Zr_2CO_2H_2$, $BP/Zr_3C_2F_2$, and $BP/Zr_3C_2O_2H_2$. High tunneling barriers appear in the vdW gap due to the weak BP-MXene interaction. A type-II band alignment is found for the $BP/Zr_2CO_2$ heterostructure, and a vertical electric field can effectively modulate the band structure of this heterostructure, and could be used to improve its performance in photoelectric devices.

## 2 Computational Details

All calculations were based on the density functional theory (DFT) scheme implemented in the Vienna ab initio simulation package (VASP) code [26, 27]. The electron-ion interactions were described with the projector-augmented wave (PAW) method [28, 29]. The generalized gradient approximation in the Perdew-Burke-Ernzerhof form (GGA-PBE) was used to describe the exchange-correlation interactions [30], together with the optB88-vdW correction for vdW interactions [31, 32]. A plane-wave basis set with an energy cutoff of 500 eV was adopted. All structures were geometry-optimized until energy and force differences were converged to $10^{-5}$ eV and 0.01 eV/Å, respectively. A vacuum layer thicker than 15 Å was used to suppress the interactions between periodic images. To minimize the lattice mismatch between BP and $Zr_{n+1}C_nT_2$ (T = O, F, OH; n = 1, 2), a 1 × 5 BP unit cell was used to match a 1 × 4 $Zr_{n+1}C_nT_2$ cell. The resulting mismatch was less than 1% (Table 1) and the averaged lattice constant was used for the corresponding heterostructure.

## 3  Results and Discussion

The optimized structures of the present BP/MXene systems are shown in Fig. 1. The BP layer in $BP/Zr_2CO_2$, $BP/Zr_2CF_2$, $BP/Zr_3C_2O_2$, and $BP/Zr_3C_2F_2$ shows a significantly distorted geometry, whereas $BP/Zr_2CO_2H_2$ and $BP/Zr_3C_2O_2H_2$ exhibit flat BP layers. The average interlayer distances are listed in Table 1. The binding energy is defined as $E_b = [E_{BP/MXene} - (E_{BP} + E_{MXene})]/N$, where $E_{BP/MXene}$, $E_{BP}$, and $E_{MXene}$ are the total energies of the BP/MXene heterostructure, BP, and MXene, respectively, and $N$ is the number of phosphorus atoms in the lower BP layer. Table 1 shows that $BP/Zr_2CO_2$, $BP/Zr_2CF_2$, $BP/Zr_3C_2O_2$, and $BP/Zr_3C_2F_2$ have very similar average interlayer distances, as well as weak binding energies, between -0.14 and -0.17 eV. $BP/Zr_2CO_2H_2$ and $BP/Zr_3C_2O_2H_2$ display smaller interlayer distances and larger binding energies, ranging from -0.38 to -0.35 eV, which indicates higher stability.

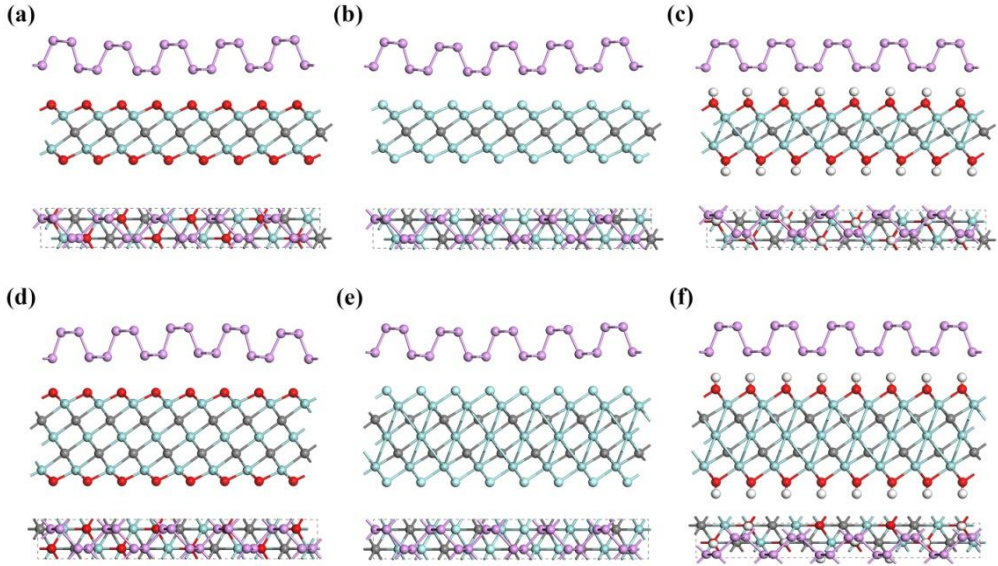

**Fig. 1** Side and top views of the optimized **(a)** BP/$Zr_2CO_2$, **(b)** BP/$Zr_2CF_2$, **(c)** BP/$Zr_2CO_2H_2$, **(d)** BP/$Zr_3C_2O_2$, **(e)** BP/$Zr_3C_2F_2$, and **(f)** BP/$Zr_3C_2O_2H_2$ heterostructures.

As indicated by the electron localization functions shown in Fig. S1, electrons are localized within the atomic layers rather than in the interlayer regions, reflecting the weak interaction between BP and MXene. The interlayer interaction can be further analyzed by examining the charge density differences, defined as $\Delta\rho = \rho(BP/MXene) - \rho(BP) - \rho(MXene)$, with $\rho(BP/MXene)$, $\rho(BP)$, and $\rho(MXene)$ being the charge densities of BP/MXene, BP, and MXene, respectively. The larger charge transfer observed in BP/$Zr_2CO_2H_2$ and BP/$Zr_3C_2O_2H_2$ (Fig. S1) denotes a stronger interlayer interaction compared to the other systems, which is consistent with the trend of the binding energies. The formation of an interface dipole is also observed, as a result of the asymmetric charge accumulation/depletion at the interface.

The band structures of BP adsorbed on $Zr_2CO_2$, $Zr_2CF_2$, $Zr_2CO_2H_2$, $Zr_3C_2O_2$, $Zr_3C_2F_2$, and $Zr_3C_2O_2H_2$ are shown in Fig. 2, with red dotted curves denoting bands dominated by BP. Since the interlayer interaction is weak, the band structure of BP remains intact upon adsorption and can thus be clearly identified. All systems except BP/$Zr_2CO_2$ are metallic. This is consistent with the fact that all surface-functionalized $Zr_{n+1}C_n$ (n = 1, 2) are metallic, except $Zr_2CO_2$. As shown in Fig. S2, $Zr_2CO_2$ is a semiconductor with an indirect band gap of 0.97 eV.

The BP/$Zr_2CO_2$ composite system has a gap of 0.63 eV, smaller than those of BP and $Zr_2CO_2$. In the case of BP/$Zr_2CF_2$, BP/$Zr_2CO_2H_2$, BP/$Zr_3C_2F_2$, and BP/$Zr_3C_2O_2H_2$, the Fermi level crosses the conduction band minimum (CBM) of BP. Since the band structure of the composite system can be described as a simple combination of the bands originating from BP and MXene, the CBM of BP can be easily identified. In the case of BP/$Zr_3C_2O_2$, the Fermi level crosses the valence band maximum (VBM) of BP.

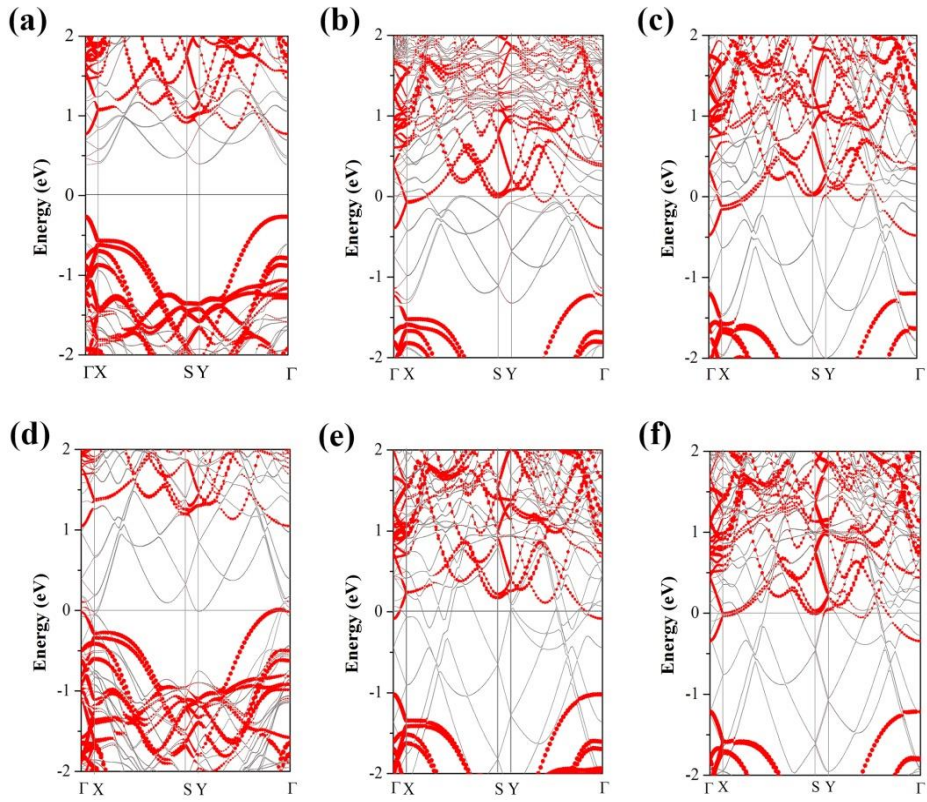

**Fig. 2** Band structures of **(a)** BP/$Zr_2CO_2$, **(b)** BP/$Zr_2CF_2$, **(c)** BP/$Zr_2CO_2H_2$, **(d)** BP/$Zr_3C_2O_2$, **(e)** BP/$Zr_3C_2F_2$, and **(f)** BP/$Zr_3C_2O_2H_2$. Red dotted curves represent contributions from BP. The Fermi energy is set to zero.

**Table 1** Calculated BP/MXene properties, including average lattice mismatch ($\varepsilon$), average interlayer distance ($d$), binding energy ($E_b$), work function of MXene ($W_{MXene}$) and BP/MXene ($W_{BP/MXene}$) structures, tunneling barrier height ($\Delta V$), width ($\omega$), and tunneling probability ($T_B$).

| | BP/$Zr_2CO_2$ | BP/$Zr_2CF_2$ | BP/$Zr_2CO_2H_2$ | BP/$Zr_3C_2O_2$ | BP/$Zr_3C_2F_2$ | BP/$Zr_3C_2O_2H_2$ |
|---|---|---|---|---|---|---|

| ε (%) | 0.27 | 0.24 | 0.27 | 0.26 | 0.37 | 0.37 |
|---|---|---|---|---|---|---|
| d (Å) | 3.04 | 3.01 | 2.2 | 3.03 | 3.03 | 2.19 |
| $E_b$ (eV) | -0.16 | -0.14 | -0.38 | -0.17 | -0.14 | -0.35 |
| $W_{MXenes}$ (eV) | 5.07 | 3.39 | 2.10 | 5.13 | 3.65 | 2.04 |
| $W_{BP/MXenes}$ (eV) | ___ | 4.30 | 4.26 | 5.25 | 5.02 | 4.18 |
| ΔV (eV) | ___ | 3.69 | 2.02 | 4.57 | 4.09 | 2.30 |
| ω (Å) | ___ | 1.28 | 0.84 | 1.36 | 1.41 | 1.0 |
| $T_B$ (%) | ___ | 8 | 29 | 5 | 5 | 21 |

In the case of BP/MXene systems with metallic MXenes, a transistor can be constructed using the MXene compound as the electrode and BP as the channel material. In such a structure, two barriers should be considered. As shown in Fig. 3(a), the first one is that between BP and MXene, in the vertical direction (labeled B), and the other is that between the electrode and channel, in the lateral direction (labeled D). The vertical Schottky barrier is defined as the energy difference between the Fermi level and the band edge of BP in the BP/MXene system [33–37]. Obviously, no vertical Schottky barrier is present for any of the five metallic systems studied here. For those systems with Fermi level crossing the conduction band (CB) of BP, the transferred electrons will not experience a Schottky barrier, whereas no Schottky barrier for hole transfer will be present for systems with Fermi level crossing the valence band (VB) of BP.

These results can be understood via a work function analysis. Due to the surface dipole effect, the work functions of MXenes are strongly dependent on the surface terminations [23, 38]. O-terminated MXenes show high work functions (even higher than Pt), while OH-terminated MXenes have ultralow work functions. Therefore, *n*-type and *p*-type Ohmic contacts can be formed in BP/MXene-based transistors with OH-terminated and O-terminated MXene electrodes, respectively. The work functions of F-terminated MXenes depend on the type of material. In the case of $Zr_2CF_2$ and $Zr_3C_2F_2$, the calculated work functions are 3.39 eV and 3.65 eV, respectively, much smaller than that of BP (4.75 eV). Therefore, *n*-type Ohmic contacts are formed in BP/$Zr_2CF_2$ and BP/$Zr_3C_2F_2$.

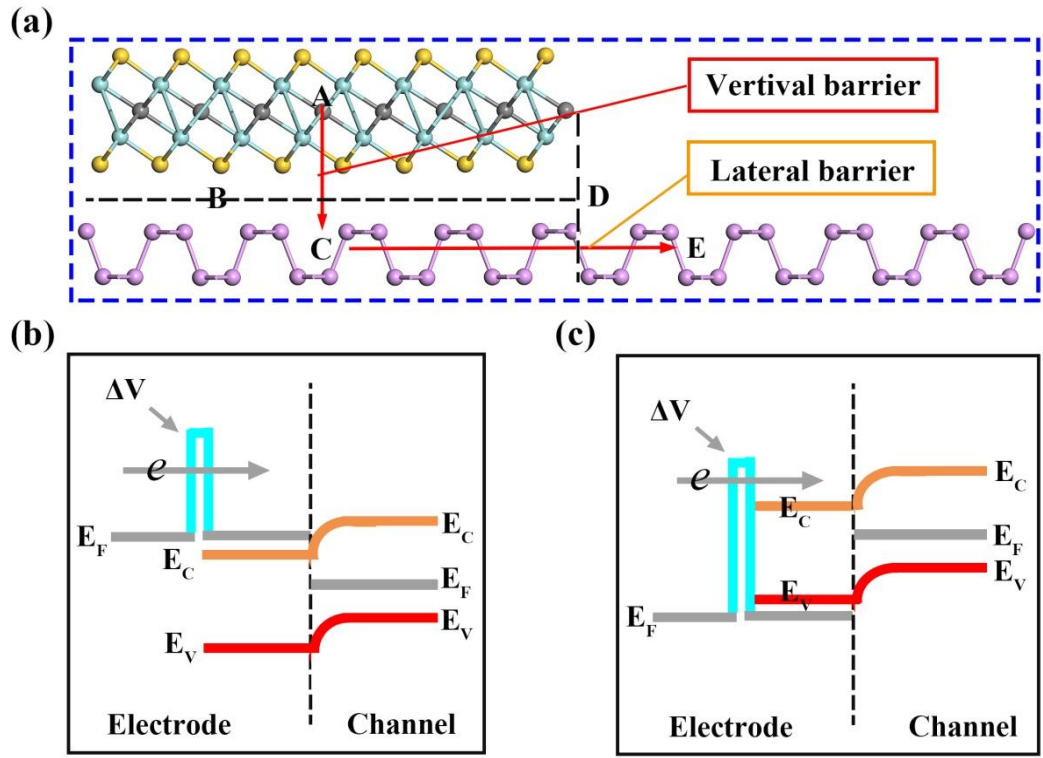

**Fig. 3 (a)** Schematic diagram of BP/MXene contacts. Red arrows indicate the pathway of carrier injection from MXene to BP. **(b)** and **(c)** represent the different band alignments and mechanisms for carrier injection from electrode to channel.

The lateral Schottky barrier height can be estimated as the energy difference between the Fermi level of the heterostructures and the band edges of BP in the channel [39]. Fig. 4 shows the work function of the interface systems and the band edges of BP. The Fermi levels of BP/$Zr_2CF_2$, BP/$Zr_2CO_2H_2$, BP/$Zr_3C_2F_2$, and BP/$Zr_3C_2O_2H_2$ are slightly lower than the CBM of BP, revealing *n*-type contacts between electrode and channel, and their Schottky barriers are 0.18, 0.14, 0.14, 0.06 eV, respectively. However, the Fermi level of the BP/$Zr_3C_2O_2$ heterostructure is lower than the VBM of the channel BP, indicating an Ohmic contact in the lateral direction.

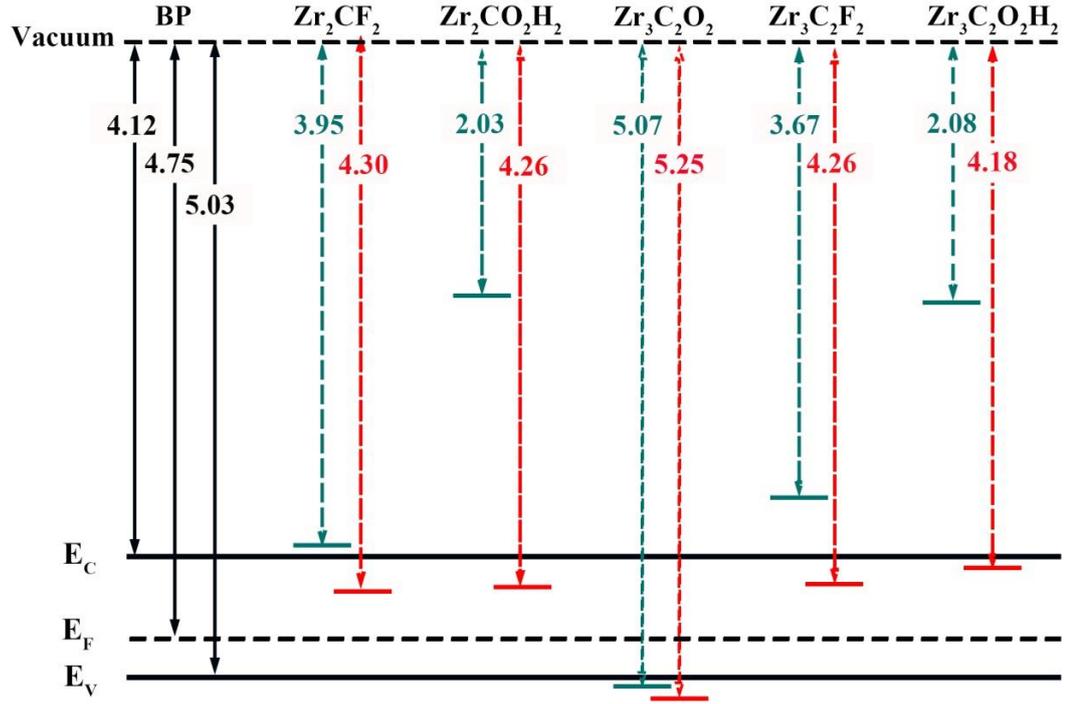

**Fig. 4** Work function alignment of pure MXene, BP, and BP/MXene interfacial systems. The green and red dashed lines represent the work functions of pure MXene and BP/MXene heterostructures, respectively.

In addition, the tunneling barriers were evaluated by plotting the electrostatic potential across the vdW gap. As shown in Fig. S3, the tunneling barrier is given by the electrostatic potential energy above the Fermi level of the BP/MXene system [40, 41]. Large tunneling barriers of 3.69, 4.57, and 4.09 eV were found for BP/$Zr_2CF_2$, BP/$Zr_3C_2O_2$, and BP/$Zr_3C_2F_2$, respectively. The tunneling barriers are lower for BP/$Zr_2CO_2H_2$ and BP/$Zr_3C_2O_2H_2$, due to the stronger interactions and more significant charge transfer in these structures. These barriers are still higher than those found in traditional BP/metal interfaces [4] because of the weak vdW interaction between BP and MXene. A square potential is assumed to predict the tunneling probabilities $T_B$, which can be defined as : $T_B = \exp(-2 \times \frac{\sqrt{2m\Delta V}}{\hbar} \times \omega_B)$, where $m$ is the free electron mass, $\hbar$ is the reduced Planck constant, $\Delta V$ is the tunneling barrier height, and $\omega_B$ represents the full width at half maximum of the tunneling barrier [41]. The highest tunneling probability is calculated to be 29% for BP/$Zr_2CO_2H_2$.

The characteristics of the BP/MXene contacts are summarized in Fig. 3(b-c). Large tunneling barriers are found for all studied BP/MXene interfaces. Ohmic contacts are formed in the vertical direction for all BP/MXene systems. In the lateral direction, low $n$-type Schottky barriers are found in the BP/$Zr_2CF_2$, BP/$Zr_2CO_2H_2$, BP/$Zr_3C_2F_2$, and BP/$Zr_3C_2O_2H_2$ heterostructures, while an Ohmic contact is formed in the BP/$Zr_3C_2O_2$ bilayer. It should be noted that the present calculations employ the GGA scheme, instead of hybrid functionals such as the Heyd-Scuseria-Ernzerhof (HSE06). Because in a FET device the interactions between electrons are largely screened by doped carriers, GGA-PBE has proved to be a good approximation for these systems [39], whereas hybrid functionals tend to overestimate the transport gap. A test HSE calculation performed for BP gave CBM and VBM values of 3.90 and 5.50 eV (Fig. S4), which are 0.22 eV higher and 0.47 eV lower than the PBE results, respectively [42]. Even considering that the BP CBM would also increase by 0.22 eV in the composite systems, an Ohmic contact would still be present for BP/$Zr_2CF_2$, BP/$Zr_2CO_2H_2$, and BP/$Zr_3C_2O_2H_2$. This confirms the validity of the main results obtained using the PBE functional in this study.

Apart from contact systems, the BP/MXene materials can also be used as heterostructures. In BP/$Zr_2CO_2$, the CBM and VBM states are localized on $Zr_2CO_2$ and BP, respectively (Fig. 5a); therefore, BP and $Zr_2CO_2$ form a type-II heterostructure [43]. Such a band alignment is beneficial for the separation of electron-hole pairs. The electrons in BP can be easily shifted to the CB of $Zr_2CO_2$. On the contrary, the holes transfer from the VB of $Zr_2CO_2$ to BP (Fig. 5b).

Moreover, an external electric field can be used as an effective tool for electronic structure engineering. The effect of a perpendicular electric field from the BP to the $Zr_2CO_2$ layer was investigated. As shown in Fig. 5(c), the band gap changes with the external electric field, increasing linearly to 0.95 eV up to an electric field of 0.4 V/Å, and then decreasing upon further electric field increases. Reversing the direction of the electric field leads to a linear decrease in the band gap. A transition from semiconductor to metallic behavior can be expected upon further increasing the reverse field. The conduction and valence band offsets (CBO and VBO, respectively) also depend on the electric field. A negative electric field results in increased band offsets, which enhances the separation of electron-hole pairs. The band edges at various electric fields are shown in Fig. 5(d). The CBM and VBM of BP gradually decrease with increasing electric field, whereas in $Zr_2CO_2$ they show a gradual increase. The electric field has little influence on the band gap of BP and $Zr_2CO_2$. At an applied external electric field larger than 0.4 V/Å, the CBM and VBM of $Zr_2CO_2$ become higher than those of BP, resulting in an inversion of the band edge states, as also confirmed by the charge density of the CBM and VBM in Fig. S5. The band edge inversion contributes to the forced electron transfer from $Zr_2CO_2$ to BP under an external electric field, to balance the original internal electric field in BP/$Zr_2CO_2$. Therefore, BP can behave as either electron donor or acceptor under a suitable electric field.

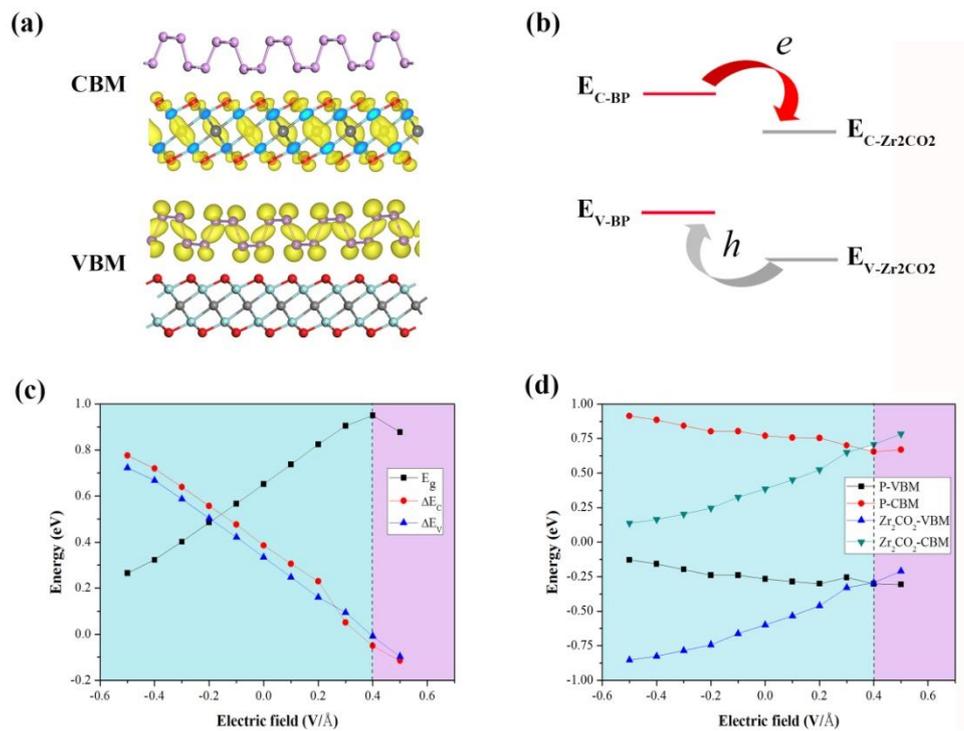

**Fig 5. (a)** Charge densities of the CBM and VBM of BP/$Zr_2CO_2$ heterostructure, plotted at an isovalue of 0.001 e/Å$^3$. **(b)** Band alignment of BP/$Zr_2CO_2$ heterostructure. **(c)** External electric

field dependence of band gap and band offsets of BP/$Zr_2CO_2$. **(d)** Band edges of BP and $Zr_2CO_2$ in BP/$Zr_2CO_2$ bilayer vs. applied electric field.

## 4 Conclusions

In summary, we have systematically investigated the interfacial properties of BP/MXene heterostructures using DFT calculations. The weak interaction between BP and MXene preserves the intrinsic properties of BP in BP/MXene-based devices. Ohmic contacts or low Schottky barriers are predicted in BP/MXene heterostructures, which are beneficial for carrier transport along the vertical and lateral directions. In addition, the semiconducting BP/$Zr_2CO_2$ bilayer shows an intrinsic type-II band alignment. An external electric field can effectively modulate the band gap and band alignment, which can be used for enhancing the performance of the BP/$Zr_2CO_2$ bilayer in photoelectric devices. The present results demonstrate that BP/MXene heterostructures provide a promising platform for various technological applications.

**Acknowledgements**

This work was partially supported by the National Natural Science Foundation of China (NSFC, 21573201), the Ministry of Science and Technology (MOST, 2016YFA0200604), and the Tianjin and Shanghai Supercomputer Centers.